\documentclass[12pt]{article}
\begin{document}
\centerline {\large \bf Better be third then second in a search for a majority
opinion}

\bigskip
\centerline{Dietrich Stauffer}

\centerline{Institute for Theoretical Physics, Cologne University}

\centerline{D-50923 K\"oln, Euroland}

\bigskip
Abstract:
Monte Carlo simulations of a Sznajd model show that if a near-consensus is 
formed out of four initially equally widespread opinions, the one which 
at intermediate times is second in the number of adherents usually loses out
against the third-placed opinion.

\bigskip
If several opinions compete against each other via mutual persuasion, and
finally a consensus or very large majority emerges, then usually (and also 
in the present work) the final winner is that opinion which at some 
intermediate stage had already a narrow majority. If at the end a tiny minority 
of dissenters remains, it seems plausible that they are the remnants of the
second-most-important opinion at some intermediate stage. However, we present
here simulations in a four-opinion model, where usually the tiny minority of
dissenters was on third and not on second place halfway through the process,
while the opinion which then was on second place finally died out. This model
is the Sznajd model on the dilute square lattice with diffusing agents of
four possible opinions.
 
The Sznajd model [1] (see [2] for a recent review) is put onto a square lattice.
Each lattice site initially is either empty, with probability 1/2, or has one
of four possible opinions 1, 2, 3, 4, with probability 1/8 each. Then at each
time step every occupied site tries to move to an empty neighbour. Afterwards
randomly selected pairs of nearest neighbours, who share the same opinion,
convince all those neighbours of the pair's opinion, which differ by at most
one unit [3]. If this is done for each lattice site, one time step is completed,
and we start again with diffusion and convincing. In this way the rigidity of
the standard Sznajd lattices is avoided; in principle everybody can exchange 
opinions with everybody else. The opinions no longer can change if they have 
settled onto the choices 1 and 3, or 2 and 4, or 1 and 4, or if one opinion
covers everybody. 

In all ten simulations of $301 \times 301$ sites, an opinion fixed point was reached
after about 4,000 to  100,000 time steps. In one case, only opinion 3 survived;
in all other cases, of the two opinions which survived at the end, one had only 
12 to 332 adherents compared with the about 45,000 of the winner. But this tiny
surviving minority was on third place half way through the process, while the
opinion which at half time had much more (4676 to about 22500) adherents 
finally had none. (In one case the leading and second opinion at half time had
about the same number of votes. The opinion ranked fourth at half time always 
died out.) So to be first or third is good, while the second place is dangerous.
(With 10000 samples of $101 \times 101$ sites, an opinion fixed point was always
reached, and the at half time second-ranked opinion  finally vanished in about
92 percent of the cases; for $31 \times 31$ exception were less rare.) 

The explanation is based on the discreteness of the four opinions and the fact
that opinion 1 is not regarded as similar to opinion 4. The two extreme opinions
1 and 4 thus can convince only one neighbouring opinion each, while the two
centrist opinions 2 and 3 have two neighbours each. After some time, most of the
opinions will be centrist (2 or 3). These two centrist opinions then fight for
a clear majority, one is winning and also will convince the extremist opinion
close to it, while the other centrist opinion is losing out completely and thus
leaves its neighbouring extremist opinion untouched. This mechanism should also
work in other models like Potts spins at low temperatures, as long as the
opinions are discrete: Being second means to lose completely; being third allows
a small chance of survival.
  
In the final fixed configuration for the 4 possible states, in 10000 simulations
of $101 \times 101$ sites each, opinions 1 and 3 survived in 41 percent of the 
cases, opinions 2 and 4 in another 41 percent, while 9 percent each had only
opinion 2 or only opinion 3 surviving. For three instead of four possible 
opinions, nearly always at the end everybody shared the centrist opinion 2; with
five possible opinions, usually a small number of opinions 1 and 5 together
with a big majority for opinion 3, and without any opinions 2 and 4, survived.
For $31 \times 31$ lattices more exceptions occur.

In summary, for survival of an opinion among four choices it may be better to 
hide on third place then to be nearly the winner.

\bigskip
\parindent 0pt
[1] K. Sznajd-Weron and J. Sznajd, Int. J. Mod. Phys. C 11, 1157 (2000).

[2]  D. Stauffer, Journal of Artificial Societies and Social Simulation 5, No.1,
paper 4 (2002) (jasss.soc.surrey.ac.uk).

[3]  G. Deffuant, D. Neau, F. Amblard and G. Weisbuch, Adv. Complex Syst. 3, 87 
(2000); R. Hegselmann and M. Krause, for Journal of Artificial Societies and 
Social Simulation 5 (2002).

\end{document}